\documentclass[12pt]{article}

\usepackage{hyperref}

\usepackage{amsmath} 
\usepackage{amssymb}
\usepackage{amsfonts}
\usepackage{amsthm}
\usepackage{graphicx}

\newtheorem{lemma}{Lemma}

\newtheorem{proposition}{Proposition}

\theoremstyle{definition}

\newtheorem{definition}{Definition}

\begin{document}
	\begin{center}
		\Large
	\textbf{One-particle approximation as a simple playground for irreversible quantum evolution}\footnote{This work is supported by the Russian Science Foundation under grant~17-71-20154.}	
	
		\large 
		\textbf{A.E. Teretenkov}\footnote{Department of Mathematical Methods for Quantum Technologies,\\Steklov Mathematical Institute of Russian Academy of Sciences,
			ul. Gubkina 8, Moscow 119991, Russia\\ E-mail:\href{mailto:taemsu@mail.ru}{taemsu@mail.ru}}
		\end{center}
		
			\footnotesize
			Both quantum information features and irreversible quantum evolution of the models arising in physical systems in one-particle approximation are discussed. It is shown that the calculation of the reduced density matrix and entanglement analysis are considerably simplified in this case. The irreversible quantum evolution described by Gorini--Kossakowski--Sudarshan--Lindblad equations in the one-particle approximation  could be defined by a solution of a Shroedinger equation with a dissipative generator. It simplifies the solution of the initial equation on the one side and gives a physical interpretation of such a Shroedinger equation with non-Hermitian Hamiltonian on the other side.
			\normalsize
			
\section{Introduction}

\noindent In works \cite{Chernega13,Chernega14,Chernega14a,Manko18} it was shown that the idea of injection of a Hilbert space into a tensor product of auxiliary  Hilbert spaces could be fruitful for the quantum information purposes. In works \cite{Teret19a, Teret19b} we showed that there is a very special case of such an injection which we called one-particle second quantization. 

There are two reasons, why such an injection is special. First of all it arises naturally when we consider one-particle approximation of both closed and open quantum systems. Namely, in \cite{Teret19a} we showed that a variety of exactly solvable quantum models which we had found in literature are actually based on one-particle approximation. Moreover, different models, for example the dissipative Jaynes--Cummings model \cite{Sachdev84} and a model of resonance two-level system decay with non-Hermitian Hamiltonian  \cite{Garraway97}, are reduced to the solution of the same equations, which becomes obvious from the one-particle second quantization point of view.

The second reason, why we are interested in such a special injection is the following one. It is possible to inject not only the states as it was done in \cite{Chernega13,Chernega14,Chernega14a,Manko18}, but also both unitary and irreversible quantum evolution, which also attracts interest in recent works \cite{Manko19}. By irreversible quantum evolution we mean the evolution according to the \textit{Gorini--Kossakowski--Sudarshan--Lindblad} (GKSL) equation \cite{Gorini76,Lindblad76} which is widely used in modern physics \cite{Accardi02, Alicki07, Breuer02, Trushechkin17, Trushechkin18, Trushechkin19}. In \cite{Teret19a} we showed there is a very natural class of GKSL equations such that their solutions after one-particle second quantization also satisfy  GKSL equations. 

Thus, it is interesting to systematically study both quantum information and dynamical features of one-particle second quantization. This is the main aim of this article.

In Sec.~\ref{sec:oneExSecQuant} we introduce the one-particle second quantization and some related definitions, then we describe quantum information theory for one-particle density matrices. Namely, in Subsec.~\ref{subsec:traces} we define the trace with respect to a set of indices and analyze the entanglement of the one-particle density matrices. Let us emphasize that results of this Subsec.~\ref{subsec:traces} show that both the calculation of partial traces and entanglement conditions are simplified a lot for the one-particle case. This simplification allows us in Subsec.~\ref{subsec:mutuial} to suggest an interpretation of terms participating in the mutual information of the one-particle density matrix. We also show that even the classical terms could be useful in discussion of the behavior of mutual information of some real physical systems. 

In Sec.~\ref{sec:GKSL} we show that the zero-temperature GKSL equation for one-particle case is reduced in a certain scene to the Liouville-von Neumann equation and the Shroedinger equations with dissipative generators. In Subsec.~\ref{subsec:GKSLquad} this result is applied to obtain one-particle solutions of zero-temperature GKSL equations with generators which are quadratic in bosonic or fermionic creation and annihilation operators. Such generators have wide applications in physics \cite{DodMan86, dodonov1985quantum, Heinosaari10, Prosen10a, Teret16, Teret17,Teret17f, Teret19x, Teret19y}. It is shown that the Gaussian solutions of such GKSL equations could be expressed in terms of solutions of the Shroedinger equations with dissipative generators.

In Conclusions we sum up our results and suggest the directions for the future study. 

\section{One-particle second quantization and its quantum information features}\label{sec:oneExSecQuant}

\noindent 	Let us consider a finite dimensional Hilbert space with a distinguished one-dimensional subspace, i.e. of the form $   \mathbb{C} \oplus \mathbb{C}^{n}$. And let $  |l \rangle, l = 0, \ldots, n $ be an orthonormal basis in it, where $ |0 \rangle  $ is a vector from the  distinguished one-dimensional subspace.  The following definition plays the central role in our work.

\begin{definition}\label{def:secQuant}
	Let $ \mathcal{X}_i $ be auxiliary Hilbert spaces with $ \dim \mathcal{X}_i \geqslant 2$. Let us introduce the linear injection $\hat{} :  \mathbb{C} \oplus \mathbb{C}^{n}\rightarrow \otimes_{i=1}^{n} \mathcal{X}_i $ which is defined on the basis by the following rule
	\begin{equation}\label{eq:OneExSecQuant}
	\begin{aligned}
	| l \rangle  &\rightarrow | \hat{l} \rangle  = | 0 \rangle_1 \otimes \cdots \otimes| 0 \rangle_{l-1} \otimes |1 \rangle_{l} \otimes | 0 \rangle_{l+1} \otimes \cdots \otimes | 0 \rangle_{n} , \qquad l \neq 0; \\
	| 0 \rangle  &\rightarrow | \hat{0} \rangle  = | 0 \rangle_1  \otimes \cdots \otimes | 0 \rangle_{n},
	\end{aligned}
	\end{equation}
	where $ | 0 \rangle_i  $ and $ | 1 \rangle_i $ are an orthonormal pair of vectors in $ \mathcal{X}_i $. We call such an injection $ \hat{} $ \textit{the one-particle second quantization}.
\end{definition}

From the physical point of view this injection allows one to regard the initial Hilbert space $  \mathbb{C} \oplus \mathbb{C}^{n}$ as a direct sum of 0-particle and 1-particle subspace of some multi-particle system. Due to linearity of such an injection for any pure state $  | \phi \rangle \in \mathbb{C} \oplus \mathbb{C}^{n} $ one has
\begin{equation*}
| \hat{\phi} \rangle = \sum_{l=0}^n \langle l | \phi \rangle | \hat{l} \rangle .
\end{equation*}
Similarly, one could define the one-particle second quantization for a density matrix $ \rho  $ in Hilbert space $   \mathbb{C} \oplus \mathbb{C}^{n}  $. (We call a matrix $ \rho  $ in the Hilbert space $   \mathbb{C} \oplus \mathbb{C}^{n}  $ a density matrix if and only if $  \rho = \rho^{\dagger} $, $ \rho $ is non-negative definite and $ \mathrm{Tr}\;\rho =1 $.) Namely, we denote
\begin{equation}\label{eq:secQuantDenMat}
\hat{\rho} = \sum_{l=0}^n \sum_{k=0}^n \langle l | \rho | k \rangle | \hat{l} \rangle \langle \hat{k} |.
\end{equation}

In the next sections we need an explicit block representation of a density matrix in  $  \mathbb{C} \oplus \mathbb{C}^{n}$
\begin{equation}\label{eq:oneExitonDM}
\rho = 
\begin{pmatrix}
\rho_{00} & \langle \psi|\\
| \psi \rangle & R
\end{pmatrix}, \qquad \rho_{00} = 1 - \mathrm{Tr} \; R, \quad R = R^{\dagger}.
\end{equation}

\begin{definition}
	If a density matrix in $  \mathbb{C} \oplus \mathbb{C}^{n}$ has form \eqref{eq:oneExitonDM} such that $ | \psi \rangle = 0 $ and $ \rho_{00} = 0$, then we call such a density matrix a \textit{strictly one-particle density matrix}. Similarly, we call a pure state $  | \phi \rangle \in \mathbb{C} \oplus \mathbb{C}^{n} $ such that $ \langle 0 | \phi \rangle = 0$ a \textit{strictly one-particle pure state}.
\end{definition}
Otherwise, we call an arbitrary density matrix in $  \mathbb{C} \oplus \mathbb{C}^{n}$  a one-particle density matrix. Of course, one could say just about a density matrix in $ \mathbb{C}^{n+1}$, but the term  ''one-particle density matrix'' supposes that a one-dimensional linear subspace is distinguished in the space in which this matrix is defined. 

Let us note that for a strictly one-particle density matrix the matrix $ R $ in  representation \eqref{eq:oneExitonDM} is a density matrix in $ \mathbb{C}^{n} $ and $ \rho = 0 \oplus R $. 

\subsection{Traces with respect to indices and entanglement}
\label{subsec:traces}

\noindent The partial trace operation for the density matrices is important both for open quantum systems and quantum information theories. The following proposition (see \cite{Teret19a}, proposition 3) shows that it takes a very simple form for the matrices obtained by one-particle second quantization.   

\begin{proposition}\label{prop:trace}
	Let $ \hat{\rho} = \sum_{l,k =0}^n \rho_{lk}|\hat{l}\rangle \langle \hat{k}| $, then the partial trace with respect to spaces indexed by $ I $ could be calculated by the formula:
	\begin{equation}\label{eq:trace}
	\mathrm{Tr}_{ \mathcal{X}_i, i \in I} \hat{\rho} = \hat{\rho}_{\overline{I}} + \left( \sum_{l \in I} \rho_{ll}\right) | \hat{0} \rangle \langle \hat{0} |, \quad \hat{\rho}_{\overline{I}} = \sum_{l,k \in \overline{I}} \rho_{lk}|\hat{l}\rangle \langle \hat{k}|,
	\end{equation}
	where $  \overline{I} = \{0, \ldots, n\} \setminus I$.
\end{proposition}

We want to work with matrices $ \rho= \sum_{l,k =0}^n \rho_{lk}|l\rangle \langle k| $ in $  \mathbb{C} \oplus \mathbb{C}^{n}$ rather than their one-particle second quantizations. This inspires the following definition.

\begin{definition}
	Let $ \rho $ be the density matrix in $  \mathbb{C} \oplus \mathbb{C}^{n}$, then \textit{the trace with respect to a set of indices} $ I \subset \{1, \ldots, n\} $ is defined by the formula
	\begin{equation}\label{eq:traceI}
	\mathrm{Tr}_I \rho = \rho_{\overline{I}} + \left(\sum_{l \in I} \rho_{ll}\right) | 0 \rangle \langle 0 |, \quad \rho_{\overline{I}} = \sum_{l,k\in \overline{I}} \rho_{lk }|l\rangle \langle k|.
	\end{equation}
\end{definition}

In representation \eqref{eq:oneExitonDM} formula \eqref{eq:traceI} takes the form 
\begin{equation}
\mathrm{Tr}_I\rho = 
\begin{pmatrix}
1 - \mathrm{Tr} \; P_I R P_I & \langle \psi| P_I\\
P_I | \psi \rangle & P_I R P_I
\end{pmatrix},
\end{equation}
where $ P_I: \mathbb{C}^n \rightarrow \mathbb{C}^{|I|}$ is projection into subspace $ \mathrm{span}\{ |k \rangle, k \in I\}$.

For example, let us apply this formula to analyze the separability of the states $ \hat{\rho} $, where $ \rho $ is a one particle density matrix, as states in the tensor product $ \mathcal{H}_1 \otimes \mathcal{H}_2 $, where $ \mathcal{H}_1 = \otimes_{i \in I} \mathcal{X}_i  $ and $ \mathcal{H}_2 = \otimes_{i \in \{1, \ldots, n\} \setminus I} \mathcal{X}_i  $. First of all, let us calculate the partial trace of a pure state $ | \phi \rangle = \sqrt{\rho_{00}} \oplus |\varphi \rangle $ (without loss of generality as the wave vector is defined up to a phase; here we assume $ \sqrt{\rho_{00}} \geqslant 0 $):
\begin{equation*}
\mathrm{Tr}_I  | \phi \rangle \langle \phi|=
\mathrm{Tr}_I \begin{pmatrix}
\rho_{00} & \sqrt{\rho_{00}} \langle \varphi|\\
\sqrt{\rho_{00}} | \varphi \rangle & | \varphi \rangle \langle \varphi|
\end{pmatrix} =
\begin{pmatrix}
1 - || P_I \varphi||^2 & \sqrt{\rho_{00}} \langle \varphi| P_I\\
\sqrt{\rho_{00}}  P_I | \varphi \rangle & P_I| \varphi \rangle \langle \varphi| P_I 
\end{pmatrix}
= |\phi_I \rangle \langle \phi_I| + || (I - P_I) | \varphi \rangle||^2 |0\rangle \langle 0|,
\end{equation*}
where $ |\phi_I \rangle = (1 \oplus P_I) | \phi \rangle $. Thus, the Schmidt number of $ |\hat{\phi} \rangle $ takes only two values: $ 1 $ or $ 2 $. A pure state is entangled if and only if its  Schmidt number is strictly greater than 1  \cite[Sec.~2.5]{Nielsen00}, \cite[Subsec~3.1.3]{Holevo12}. Hence, we have proved the following proposition.

\begin{proposition}
	$ |\hat{\phi} \rangle $, where   $ | \phi \rangle = \phi_0 \oplus |\varphi \rangle $, is entangled if and only if  $  P_I | \varphi \rangle \neq 0$ and $  P_{\{1, \ldots,n \} \setminus I} | \varphi \rangle \neq 0$, i.e. if and only if the support of $ | \phi \rangle $ contains at least one index from $ I $ and at least one (strictly) positive index outside $ I $.
\end{proposition}

The tensor product of two density matrices $ \hat{\rho}_{1} \otimes \hat{\rho}_{2} $, where $ \rho_{1} $ and $ \rho_{2} $ are one-particle density matrices, does not necessarily have the form $ \hat{\rho}_{12} $, where $ \rho_{12} $ is some one-particle density matrix.  Actually, this is the case if and only if $ \hat{\rho}_{1} $ or $ \hat{\rho}_{2} $ is the vacuum $ | \hat{0} \rangle \langle \hat{0}| $. Hence, any separable state takes the form
\begin{equation*}
\rho =
p_1 \rho_1 \otimes |0\rangle \langle 0| + p_2 |0\rangle \langle 0| \otimes \rho_2 =
\begin{pmatrix}
\rho_{00} & p_1 \langle \psi_1 | \oplus p_2\langle \psi_1 | \\
p_1 |\psi_1 \rangle \oplus p_2|\psi_1 \rangle & p_1 R_1 \oplus p_2 R_2,
\end{pmatrix}
\end{equation*}
where $ p_1, p_2 \geqslant 0 $, $ p_1 + p_2 =1 $, i.e. a convex combination of one-particle product states. Hence, for an arbitrary one-particle  separable state one has $  P_I R P_{\{1, \ldots,n \} \setminus I} = 0  $ and it is necessary for separability of $ \hat{\rho} $. For a strictly one-particle density matrix $ \rho=0 \oplus R $, it is also sufficient for separability of $ \hat{\rho} $. It follows from the fact that one could assume
\begin{equation*}
p_1 = \mathrm{Tr}\; P_I R P_I, \quad R_1 = \frac{1}{p_1} P_I R P_I, \quad  p_2 = \mathrm{Tr}\;  P_{\{1, \ldots,n \} \setminus I} R  P_{\{1, \ldots,n \} \setminus I}, \quad R_2 = \frac{1}{p_2}  P_{\{1, \ldots,n \} \setminus I} R  P_{\{1, \ldots,n \} \setminus I},
\end{equation*}
if $ p_1 p_2 \neq 0 $. If $ p_1 p_2 = 0 $, then  $ \hat{\rho} $ becomes a product state. Thus, we have proved the following proposition.
\begin{proposition}
	For an arbitrary one-particle  state $ \rho $ in representation \eqref{eq:oneExitonDM} the condition $  P_I R P_{\{1, \ldots,n \} \setminus I} = 0  $ is necessary for separability of $ \hat{\rho} $.  For a strictly one-particle density matrix, it is also sufficient for separability of $ \hat{\rho} $.
\end{proposition}

A strictly one-particle state is described by a density matrix $ R $ in $ \mathbb{C}^n $. In such a case the trace with respect to the set of indices $ I $ is fully described by the quantum operation $ P_I \cdot P_I:\mathbb{C}^{n \times n} \rightarrow \mathbb{C}^{|I| \times |I|} $ (the quantum operation is a completely positive trace non-increasing map \cite[Subsec.~8.2.4]{Nielsen00}) defined by the formula $ P_I \cdot P_I (\rho) = P_I \rho P_I $. This suggests the following generalization: Let  $ \Phi_I:\mathbb{C}^{n \times n} \rightarrow \mathbb{C}^{|I| \times |I|} $ be an arbitrary quantum operation, then one could define for any strictly one-particle matrix $ 0 \oplus R $ a generalized reduced density matrix $ \rho_{\Phi_I} $ by the formula $ \rho_{\Phi_I}  = (1 - \mathrm{Tr}\;\Phi_I(R)) \oplus \Phi_I(R)$. From the physical point of view the case, when the Kraus operators of the operation $ \Phi_I $ are not orthogonal projectors,  such an operation is interpreted as non-ideal measurement. So $ \rho_{\Phi_I} $ could be interpreted as a density matrix of a subsystem which is not ideally separated from the whole system.  Each site $ |i \rangle \langle i|, i=1, \ldots, n $ is contained in such a system with probability $ \mathrm{Tr}\; \Phi_I(|i \rangle \langle i|) $ which can be different from zero and one. At the same time for $ \Phi_I = P_I \cdot P_I $ one obtains $  \mathrm{Tr}\; \Phi_I(|i \rangle \langle i|) = \mathrm{Tr}\; P_I |i \rangle \langle i| P_I $ which is $ 1 $ if $ i \in I $ and $ 0 $, otherwise.

\subsection{Mutual information}\label{subsec:mutuial}

\noindent The correlations shared between two subsystems could be measured by mutual information \cite{Volovich11, Arefeva16} sometimes called information correlation \cite[Sec.~7.5]{Holevo12}. To deal with it in our case we need the following lemma. 

\begin{lemma}\label{lem:fMutInf}
	Let $ \rho $ be a density matrix in $  \mathbb{C} \oplus \mathbb{C}^n  $ with $ | \psi \rangle  = 0 $ in representation \eqref{eq:oneExitonDM} and $ I_1 \sqcup I_2 = \{1, \ldots, n\} $, then for any function $ f $ defined on interval $ [0,1] $ one has
	\begin{equation}\label{eq:auxFormula}
	\mathrm{Tr} f(\mathrm{Tr}_{I_1} \rho) + \mathrm{Tr} f(\mathrm{Tr}_{I_2} \rho) - \mathrm{Tr} f( \rho) = 
	$$
	$$
	\mathrm{Tr} f(P_{I_1} R P_{I_1}) + \mathrm{Tr}f(P_{I_2} R P_{I_2}) - \mathrm{Tr}f(R)+ 
	$$
	$$
	f( \rho_{00} + \mathrm{Tr} P_{I_1} R P_{I_1}) + f(\rho_{00} + \mathrm{Tr} P_{I_2} R P_{I_2}) - f( \rho_{00}),
	\end{equation}
	where $ \rho_{00} $ and $ R $ are defined by representation \eqref{eq:oneExitonDM} of $ \rho $.
\end{lemma}
(As the function $ f $ in this lemma is applied only to Hermitian matrices it needs to be defined only on its spectrum. Namely, if $ M $ is a Hermitian matrix, then it could be represented in the form $ M = U \mathrm{diag} \{\lambda_1, \cdots, \lambda_n\} U^{\dagger} $, where $ U $ is a unitary matrix and $ \mathrm{diag} \{\lambda_1, \cdots, \lambda_n\} $ is a diagonal matrix with $ \lambda_i $ on its diagonal, then $ f(M) $ could be defined by the formula $ f(M) = U  \mathrm{diag} \{f(\lambda_1), \cdots, f(\lambda_n)\} U^{\dagger} $.)

The proof of this lemma is based on straightforward calculation of reduced density matrices $ \mathrm{Tr}_{I_1} \rho $ and $ \mathrm{Tr}_{I_2} \rho $ by formula \eqref{eq:traceI}. If one chooses 
\begin{equation}\label{eq:fEntorpy}
f(x) = 
\begin{cases}
0, \qquad &x = 0,\\
-x \ln x, \qquad & 0 <x \leqslant 1,
\end{cases}
\end{equation}
the left-hand side of \eqref{eq:auxFormula} becomes the definition of the mutual information $ \mathcal{I}_{I_1 I_2}(\hat{\rho}) $ between the subsystems $ I_1 $ and $ I_2 $ \cite[Sec.~7.5]{Holevo12}.

Let us define a completely positive trace-preserving map $ \Phi = \Phi_{I_1} + \Phi_{I_2} $, where $ \Phi_{I} = P_I \cdot P_I $, and a classical probability distribution on three elementary events $ \pi = (p_0,p_1, p_2) $, where $ p_0 = \rho_{00} $, $ p_1  = \mathrm{Tr} \; \Phi_{I_1}(R)$, $ p_2  = \mathrm{Tr} \; \Phi_{I_2}(R) $. Let us also define the probability distributions  $ \pi_1 = (p_0  + p_2,p_1) $ and $ \pi_2 = (p_0 + p_1,p_2) $. For a density matrix one could define von Neumann entropy by the formula $ S(\rho) = \mathrm{Tr} \; f(\rho) $, where the function $ f $ is defined by formula \eqref{eq:traceI}. Similarly, one could define  Shannon entropy for a discrete classical distribution $ \pi = (p_i) $ by formula $ S_{\rm cl}( \pi) = \sum_i f(p_i) $. After all these definitions the result of calculation of the mutual information by lemma \ref{lem:fMutInf} could be represented in the following way.   

\begin{proposition}\label{prop:mutInfInt}
	Let $ \rho $ be a density matrix in $  \mathbb{C} \oplus \mathbb{C}^n  $ with $ | \psi \rangle  = 0 $ in representation \eqref{eq:oneExitonDM} and $ I_1 \sqcup I_2 = \{1, \ldots, n\} $, then
	\begin{equation*}
	\mathcal{I}_{I_1 I_2}(\hat{\rho}) = \Delta S_{\Phi}(R) + \mathcal{I}_{\rm cl}(\pi),
	\end{equation*}
	where
	\begin{align*}
	\Delta S_{\Phi}(R) &\equiv S(\Phi(R)) - S(R),\\
	\mathcal{I}_{\rm cl}(\pi) &\equiv S_{\rm cl}( \pi_1) +  S_{\rm cl}(\pi_2) - S_{\rm cl} (\pi).
	\end{align*}
\end{proposition}

This result has clear physical meaning: the term $ \Delta S_{\Phi}(R) $ describes the entropy change due to decoherence as a result of measurement in which subsystem the particle is and $ \mathcal{I}_{\rm cl}(\pi) $ is a pure classical term. In the case of strictly one-particle states, i.e. if $ \rho_{00} = 0 $, the last term is reduced to the form $ \mathcal{I}_{\rm cl}(\pi) =S_{\rm cl} (\pi)$, i.e. is defined by classical information from the measurement in which subsystem the particle is if one forgets about internal structure of these subsystems. 

Proposition \ref{prop:mutInfInt} could also be used for generalization of mutual information to an arbitrary quantum instrument $ \{\Phi_{I_1},\Phi_{I_2}\} $  with binary output, which develops the idea of non-ideally separated subsystems introduced in the previous subsection. Recall that the quantum instrument with binary output is defined by the set $ \{\Phi_{I_1},\Phi_{I_2}\} $, where $ \Phi_{I_1} $ and $ \Phi_{I_2} $ are completely positive trace non-increasing maps and $ \Phi = \Phi_{I_1} + \Phi_{I_2} $ is a completely positive trace-preserving map \cite[Sec.~4.1]{Davies76}. 

\begin{figure}[h!]
	\centering
	\includegraphics[width=.5\linewidth]{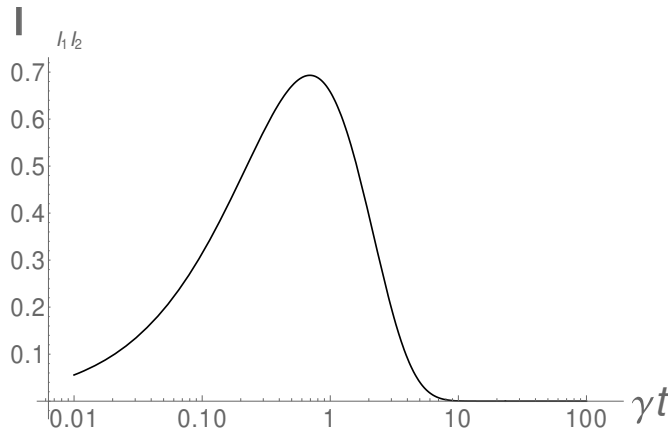}
	\caption{Mutual information time-dependence for the simplest classical Markov dynamics.}
	\label{fig1}
\end{figure}

If $ \Phi(R) = R $, i.e. there are no quantum correlations between the subsystems, then $ \mathcal{I}_{I_1 I_2}(\hat{\rho}) =\mathcal{I}_{\rm cl}(\pi)$. But in this simple case it could be useful. In \cite{Bradler10} the mutual information dynamics in the model of the Fenna-Matthews-Olson complex was simulated. In \cite{Arefeva16} the characteristic bell-shape behavior of the mutual information as a function of time was noted. In particular, both in short-time and long-time limits the mutual information tends to zero. Now we can show that such behavior can be explained by purely classical reasons. Namely, $ \mathcal{I}_{\rm cl}(\pi) = 0 $, if $ p_i=0 $ for $ i =0 $ or for $ i =1 $, i.e. when there are for sure no particles in one of subsystems. But this is the case both for the initial and final state considered in \cite{Bradler10}. Namely, the initial state was a pure strictly one-particle state $ |l \rangle, l \neq 0 $ and the final state was a convex combination of  $ |0 \rangle \langle 0 | $ and $ |s \rangle \langle s |, s \neq 0$ (from the physical point of view this state describes sink). Moreover, if we consider a decay from one subsystem to the other one, which is the simplest Markovian dynamics, i.e. if we assume $ p_0 =0 $, $ p_1 = e^{- \gamma t} $, $ p_2 = 1- e^{- \gamma t} $, then we obtain a bell-shape curve for $ \mathcal{I}_{I_1 I_2}(\hat{\rho}) $ as it is shown in Fig.~\ref{fig1}.

\section{GKSL equation}\label{sec:GKSL}

In \cite{Teret19a} it was shown that the representation \eqref{eq:oneExitonDM}  of the one-particle density matrix leads to natural connection of a finite-dimensional von Neumann equation with dissipative generators and zero-temperature GKSL master equations. Here we generalize proposition 1 from \cite{Teret19a} by the following one (which is reduced to proposition 1 from \cite{Teret19a} in the case, when $ H(t) $ and $ L_l(t) $ are time-independent and $ | \psi(t) \rangle = 0$, after taking into account a slightly different notation in \cite{Teret19a}).

\begin{proposition}\label{prop:GKSL}
	The solution of GKSL equation for the density  matrix $ \rho(t)  $ in $ \mathbb{C} \oplus \mathbb{C}^n $
	\begin{equation}\label{eq:GKSL}
	\frac{d}{dt} \rho(t) = - i [0 \oplus H(t) , \rho(t)]+ \sum_{l=1}^K \left(L_l(t) \rho(t) L_l^{\dagger}(t) - \frac12 L_l^{\dagger}(t) L_l(t) \rho(t) - \frac12 \rho(t) L_l^{\dagger}(t) L_l(t)  \right),
	\end{equation} 
	where $ H(t) \in \mathbb{C}^{n \times n} $ ($ H(t) = H^{\dagger}(t) $) and $ L_l(t) = |0 \rangle \langle f_l (t)| \in \mathbb{C}^{(n+1) \times (n+1)}$  are continious functions in $ t $ on half-line $ \mathbb{R}_+ $ ($  |f_l (t) \rangle \in \mathbb{C} \oplus \mathbb{C}^n$, $ \langle f_l (t)|0 \rangle  = 0 $), has form \eqref{eq:oneExitonDM}, where $ | \psi(t) \rangle $ and $ R(t) $ satisfy the equations
	\begin{align}
	\label{eq:psiEvol}
	\frac{d}{dt}| \psi(t) \rangle &= - A(t) | \psi(t) \rangle,\\
	\label{eq:REvol}
	\frac{d}{dt}R(t) &= - A(t) R(t) - R(t) A^{\dagger}(t),
	\end{align}
	where
	\begin{equation}\label{eq:accr}
	A(t) = \frac12 \sum_{l=1}^K |f_l (t) \rangle \langle f_l (t)| + i H(t)
	\end{equation}
	is an accretive matrix.
\end{proposition}

Recall that the matrix $ A $ is accretive if its hermitian part $ A + A^{\dagger} $ is non-negative definite \cite{George05}. For time-indpendent $ A $ by the finite-dimensional reduction of the Lumer--Phillips theorem (see \cite{Engel2000}, Th. 3.15) $ -A $ is a generator of a contraction semigroup and by the GKSL theorem \cite{Gorini76,Lindblad76} the superoperator on the left-hand side of \eqref{eq:GKSL} is a generator of a completely positive trace preserving semigroup. Hence, proposition \ref{prop:GKSL} could be regarded as an algorithm with a contraction semigroup at the input and a completely positive trace preserving semigroup at the output. 

Let us note that Eq.~\eqref{eq:GKSL} is a GKSL equation of the very special form. From the physical point of view such equation occurs (in the weak coupling limit) in the case of a zero-temperature reservoir.  

The proof of proposition \ref{prop:GKSL} is based on direct substitution of \eqref{eq:oneExitonDM} into \eqref{eq:GKSL}.

Let us note that proposition~1 from \cite{Teret19a} was based on the concept of the normalization reconstruction introduced there (see, definition 1 from \cite{Teret19a}). This concept does not take into account the vector $ | \psi \rangle  $ in  \eqref{eq:oneExitonDM} assuming it to be zero. But in \cite{Teret19b} it was shown that the evolution of this part of the density matrix gives important physical predictions and, hence, is also interesting. In this paper we have also generalized proposition~1 from \cite{Teret19a} to the time-dependent case. In spite of the fact that this is straightforward it could be useful for the problems of quantum control which achieve rising interest now \cite{Pechen17, Ilin18, Morzhin2019a, Morzhin2019b}. Eq.~\eqref{eq:psiEvol} also naturally arises in the pseudomode approach \cite{Imamoglu94, Garraway96, Garraway97a, Garraway97, Dalton01, Garraway06, Luchnikov19}.

For example, let us consider a homogenious reservoir, i.e. assume that $ |f_l (t) \rangle = \sqrt{\gamma(t)}|l \rangle $, where $ |l \rangle, l =1, \ldots, n $ is an orthonormal basis in the subspace $  \mathbb{C}^n $ of $ \mathbb{C} \oplus \mathbb{C}^n $ (orthogonal to $ |0\rangle $) and $  \gamma(t) \geqslant 0 $ is a continuous function on $ \mathbb{R}_+ $. Then Eq.~\eqref{eq:REvol} takes the form
\begin{equation*}
\frac{d}{dt}R(t)  = -  \gamma(t) \left( \sum_{l=1}^n |l \rangle \langle l | R(t) + R(t)  \sum_{l=1}^n |l \rangle \langle l | \right),
\end{equation*}
If one calculates the trace of left- and right-hand sides of this equation, then one has
\begin{equation*}
\frac{d}{dt} \mathrm{Tr} R(t) =- \gamma(t) \mathrm{Tr} R(t).
\end{equation*}
After integration one has
\begin{equation*}
\mathrm{Tr} R(t) = e^{- \int_0^t ds \gamma(s)} \mathrm{Tr} R(0).
\end{equation*}
Taking into account the normalization condition of density matrix \eqref{eq:oneExitonDM} one obtains 
\begin{equation*}
\rho_{00}(t) = 1- e^{- \int_0^t ds \gamma(s)} (1 - \rho_{00}(0)).
\end{equation*} 
From the physical point of view this is the population of the ground state. Thus, for zero temperature time-dependent homogeneous reservoir we have derived the analytical expression for this quantity. It is important, for example, for energy transfer models \cite{Kozyrev17}. 

Let us also note that in the case $ \gamma(t) = \gamma > 0 $  one has strictly contracting evolution for all ''excited'' observables and the dynamics of the observable $ | 0 \rangle \langle 0| $ is defined by the normalization condition. This is similar to the case of theorem~3 in \cite{Volovich18} and the understanding of the connection between infinite-dimensional Hilbert space of our results and the results of \cite{Volovich18} could be fruitful for future study.

The standard von Neumann equation is reduced to the  Shroedinger equation for the pure states. Similarly, we obtain the following proposition. 
\begin{proposition}
	Let $ | \psi(t) \rangle $ satisfy Eq.~\eqref{eq:psiEvol}, then the matrix $ R(t) =  | \psi(t) \rangle \langle \psi(t)|  $  satisfies Eq.~\eqref{eq:REvol}.
\end{proposition}

Actually, all dynamics could be expressed in terms of the propagator of the Shroedinger equation with a dissipative generator in the following way.
\begin{proposition}\label{prop:solV}
	Let $ V(t) $ be a solution of the Cauchy problem
	\begin{equation}\label{eq:solV}
	\begin{cases}
	\frac{d}{dt} V(t) &= - A(t) V(t), \\
	V(0) &= I,
	\end{cases}
	\end{equation}
	then the solutions of equations \eqref{eq:psiEvol} and \eqref{eq:REvol} have form
	\begin{align*}
	| \psi(t) \rangle &= V(t) | \psi(0) \rangle, \\
	R(t) &= V(t) R(0) V^{\dagger}(t).
	\end{align*}
\end{proposition}

\subsection{GKSL equations with a quadratic generator}
\label{subsec:GKSLquad}

A straightforward time-dependent generalization of proposition~2 from \cite{Teret19a} takes the following form.

\begin{proposition}\label{prop:OneExSecQuantGKSL}
	Let $ a_l $ be arbitrary operators such that $ a_l^{\dagger} | \hat{0} \rangle = | \hat{l} \rangle $ and  $ a_l | \hat{l} \rangle = |\hat{0} \rangle $. Let $ \rho(t) $ satisfy Eq.~\eqref{eq:GKSL} with coefficients defined in
	proposition \ref{prop:GKSL}. Then $ \hat{\rho}(t) $ defined by formula \eqref{eq:secQuantDenMat} satisfies the equation
	\begin{equation}\label{eq:OneExSecQuantGKSL}
	\frac{d}{dt} \hat{\rho}(t) = - i [\hat{H}(t), \hat{\rho}(t)]+ \sum_{l, k=1}^n \left(\Gamma_{kl}(t) a_l \hat{\rho}(t) a_k^{\dagger} - \frac12 \Gamma_{lk}(t) a_l^{\dagger} a_k \hat{\rho}(t) - \frac12 \Gamma_{lk}(t) \hat{\rho}(t)  a_l^{\dagger} a_k  \right),
	\end{equation} 
	where
	\begin{equation*}
	\hat{H}(t) = \sum_{l,k = 1}^n H_{lk}(t) a_l^{\dagger} a_k, \qquad \Gamma(t) =  \sum_{l=1}^K |f_l (t) \rangle \langle f_l (t)|.
	\end{equation*}
\end{proposition}

Thus, one could regard Eq.~\eqref{eq:OneExSecQuantGKSL} as one-particle second quantization of Eq.~\eqref{eq:GKSL}.

Let us apply now this proposition to the bosonic quadratic generator at zero temperature, assuming $ \mathcal{X}_i = \ell_2 $ in definition \ref{def:secQuant} and choosing $ a_l^{\dagger}, a_k $ as bosonic creation and annihilation operators satisfying canonical commutation relations $ [a_l, a_k^{\dagger}] = \delta_{lk}, [a_l, a_k] =0 $. Thus, one could obtain one-particle solutions of Eq.~\eqref{eq:OneExSecQuantGKSL}. At the same time one could obtain Gaussian solutions for general GKSL equations with bosonic quadratic generators. A Gaussian state $ \hat{\rho} $ could be fully characterized by a vector of means $ m_j \equiv \mathrm{Tr} a_j \hat{\rho} $ and matrices of second central moments $ Y_{ij} \equiv \mathrm{Tr}  a_i^{\dagger} a_j \hat{\rho}  - \overline{m}_i m_j$, $ Z_{ij} \equiv \mathrm{Tr}  a_i a_j \hat{\rho} - m_i m_j $ \cite[Sec.~12.3]{Holevo12}. Thus, it is interesting to compare finite-dimensional equations \eqref{eq:psiEvol}, \eqref{eq:REvol}  for $ | \psi(t) \rangle $, $ R(t) $ defining the one-particle solutions of \eqref{eq:OneExSecQuantGKSL} and the finite-dimensional equations for $ m(t) $, $ Y(t) $ and $ Z(t) $. For this purpose let us adapt general theorem 1 from \cite{Teret19x} to our special case in the following way.

\begin{proposition}\label{prop:bosGaussian}
	Let the solution $ \hat{\rho}(t) $  of Eq.~\eqref{eq:OneExSecQuantGKSL}, where $ a_l^{\dagger}, a_k $ are bosonic creation and annihilation operators, be a density matrix which has finite moments $ m(0) $, $ Y(0) $ and $ Z(0) $ at initial time, then it has finite moments $ m(t) $, $ Y(t) $, $ Z(t) $ for $ t>0 $ and they satisfy the equations
	\begin{align}
	\label{eq:mEvol}
	\frac{d}{dt} m(t) &= - A(t) m(t),\\
	\label{eq:YEvol}
	\frac{d}{dt} Y(t) &=  - \overline{A}(t) Y(t) - Y(t) A^{T}(t), \\
	\label{eq:ZEvol}
	\frac{d}{dt} Z(t) &=  - A(t) Z(t) - Z(t) A^T(t)
	\end{align}
	for $  t \geqslant 0 $, where $ A(t) =  i H(t) + \frac12 \Gamma(t) $, i.e. the same as in formula \eqref{eq:accr}.
\end{proposition}

Let us note that in this proposition the notation $ \hat{\rho}(t) $ means only that this is a density matrix in $ \otimes^n \ell_2$ and it does not have to be a result of one-particle second quantization.  

The vector of means $ m(t) $ satisfies the same equation as $ | \psi(t) \rangle $,   $ Y^T(t) $ satisfies the same equation as $ R(t) $, $ Z(t) $ satisfies the same equation as $  | \psi(t) \rangle (| \psi(t) \rangle)^T$. It leads to the following proposition.

\begin{proposition}
	Let $ V(t) $ be a solution of Cauchy problem \eqref{eq:solV}, then solutions of \eqref{eq:mEvol}, \eqref{eq:YEvol} and \eqref{eq:ZEvol} have the form
	\begin{align*}
	m(t) &= V(t) m(0),\\
	Y(t) &= \overline{V}(t) Y(0) V^T(t),\\
	Z(t) &= V(t) Z(0) V^{T}(t).
	\end{align*}
\end{proposition}

Thus, the Gaussian solutions of \eqref{eq:OneExSecQuantGKSL} are fully defined by one-particle propagator $ V(t) $. In the case of time independent $ A(t) $ this proposition constructs a Gaussian solution of \eqref{eq:OneExSecQuantGKSL}  by a contraction semigroup in the same way as propositions \ref{prop:GKSL}, \ref{prop:solV} and \ref{prop:OneExSecQuantGKSL} construct a one-particle solution of \eqref{eq:OneExSecQuantGKSL}. This idea could be used for further generalizations for contraction semigroups in infinite dimensional spaces. 

One could also apply proposition \ref{prop:OneExSecQuantGKSL} to the fermionic case, assuming $ \mathcal{X}_i = \mathbb{C}^2$ in definition \ref{def:secQuant} and choosing $ a_l^{\dagger}, a_k $ as creation and annihilation operators satisfying canonical anticommutation relations $ \{a_l, a_k^{\dagger}\} = \delta_{lk}, \{a_l, a_k\} =0 $. But as well as for bosons one could obtain even Gaussian solutions in this case.  For fermions the superselection rules \cite{Amosov17} lead to $  \mathrm{Tr} a_j \hat{\rho}  = 0 $, then one defines $ Y_{ij} \equiv \mathrm{Tr}  a_i^{\dagger} a_j \hat{\rho}  $ and $ Z_{ij} \equiv \mathrm{Tr}  a_i a_j \hat{\rho} $. It is interesting, that for the fermionic case the same equations for $ Y(t) $ and $ Z(t) $ are held. By theorem~2 from \cite{Teret19x} we obtain the following proposition.

\begin{proposition}
	Let the solution $ \hat{\rho}(t) $  of Eq.~\eqref{eq:OneExSecQuantGKSL}, where $ a_l^{\dagger}, a_k $ are fermionic creation and annihilation operators, be a density matrix which has finite moments $ Y(0) $ and $ Z(0) $ at initial time, then it has finite moments $ Y(t) $, $ Z(t) $ for $ t>0 $ and they satisfy equations \eqref{eq:YEvol} and \eqref{eq:ZEvol}.	
\end{proposition}

Similarly to proposition~\ref{prop:bosGaussian} the notation $ \hat{\rho}(t) $ means only that this is a density matrix in $ \otimes^n \mathbb{C}^2$ and it does not have to be a result of one-particle second quantization.

\section{Conclusions}

\noindent
In this paper, we have discussed one-particle second quantization. We have obtained the reduced density matrices for the one-particle case, which allowed us to obtain entanglement conditions and calculate the mutual information. We have also considered zero-temperature GKSL equations and their one-particle second quantizations. This has provided us with the one-pariticle solutions of zero-temperature GKSL equations with quadratic generators. We have shown that they are expressed in terms of the propagator of the Shroedinger equation with a dissipative generator. Moreover, we have shown that both for the fermionic and for the bosonic case the Gaussian solution are also expressed in terms of this propagator.

From the mathematical point of view we suggest the following directions of future development: 
\begin{enumerate}
	\item  To generalize the non-ideal subsystem separation concept introduced in Subsec.~\ref{subsec:traces} at least for (non necessarily strictly) one-prarticle states and, if it is possible, for arbitrary states.
	\item To generalize the algorithm (from Subsec.~\ref{subsec:GKSLquad}) of producing the one-particle and Gaussian solutions of GKSL equations from finite-dimensional contraction semigroup for infinite-dimensional Hilbert and Banach spaces. 
\end{enumerate}

From the physical point of view the main direction of future development consists in applying the results of this article to real and model physical systems in a manner contemplated in \cite{Teret19a,Teret19b}. Namely, by coupling the methods described here and the theory for the generalized Friedrichs model \cite{Friedrichs48,Kossakowski07,Prigogine95,Petrosky91,Antoniou93, Karpov00, Parravicini80, Antoniou03} one could obtain exact non-Markovian dynamics of a subsystem and analyze its properties, which is under active discussion in contemporary physical literature  \cite{Wilkie00, ChruscinskiKossakowski10, Trushechkin19x, Trushechkin19y, Teret19a,Teret19b}. 

\section{Acknowledgments}
The author thanks A.\,S.~Trushechkin, I.\,V.~Volovich, S.\,V.~Kozyrev, B.\,O.~Volkov  and A.\,I.~Mikhailov for the fruitful discussion of the problems considered in the work. I would also like to thank the referee for thorough reading of my work and critical remarks which allowed me to improve the text of this article significantly.

\end{document}